# To do things with words (only): An introduction to the role of "noise" in coordination dynamics without equations.

Julien Lagarde

EuroMov, UFR STAPS, University of Montpellier

Julien.lagarde@umontpellier.fr

**Abstract:** Uncertainty, spatial or temporal errors, variability, are classic themes in the study of human and animal behaviors. Several theoretical approaches[1] and concepts have been adopted to tackle those issues, often considering the CNS as an observer, using Shannon information and entropy, signal to noise ratio, and recently a Bayesian approach, and free energy minimization. In the coordination dynamics framework, addressing pattern formation processes underlying cognitive functions, and goal directed human movement among others, the tools employed originate from the statistical physics of Brownian motion and stochastic processes. The relations between those theories/ concepts have been drawn for some cases in their original fields, for example between free energy minimization and diffusion model in a force field (Jordan et al., 1998). Here an introductory presentation of the approach of "noise" in coordination dynamics is proposed, aimed at the experimentalist.

Patterns of coordination, representing highly integrated sensori-motor units, provide the biological system a massive simplification of the control problem of its very many degrees of freedom (joints, muscles, neurons). So-called collective variables have been identified empirically across a variety of tasks, showing that goal directed behavior in a very complex system can be captured by one or few quantities, but also that coordination phenomena are more the rule than the exception. These key variables characterize the degree of order, of coordination, within a functional system.

How noise can affect coordination patterns and lead to the variability observed in actual behavior? The basic message here will be that the measured behaviour is the output of the interplay between internal noise and stability. The underlying processes are considered by distinguishing determinist and stochastic components. This is the classical approach to noisy systems, or stochastic processes, started in physics and mathematics, and more recently applied to biology, coming back to Einstein treatment of the Brownian motion problem completed in 1905 (see for a presentation Gardiner, 1990). The determinist component provides the *structure, or shape,* requiring stability (see Strogatz, 2004). The stochastic component introduces the contribution of random forces. However, the observed behaviour is truly the product of the interplay of the two components, and cannot be deduced from only one of these two (see van Kampen, 2007).

Outline of the methodology

This approach requires the identification of the states of the system under scope, which entails finding the attractors, types and numbers, and instabilities or bifurcations (changes) between then (see Kelso, 1995). As already stated above, this theoretical and experimental framework focuses on the stability of coordination patterns, the definition of which is derived from the application of mathematical theory of dynamical systems to biological systems (Kelso, 1995; Strogatz, 1994). Two complementary objects of inquiry define this research program: The determiners of stability of stationary, permanent, coordination patterns, and the determiners of transients behaviours occurring during phase transitions, responsible for the formation, "dissolution", and selection of coordination patterns. One aims at understanding the processes by which multiple sensori-motor degrees of freedom are bound into integrated efficient or deficient coordination patterns, and the potential breakdown of these assemblies. Systematic relationships between effectors can reduce the number of degrees of freedom (i) to form functional units, or patterns, (ii) to shape patterns of motion in a dynamical system's state variables (e.g., position and velocity), and (iii) to shape spatiotemporal patterns of movement during skilled actions (Saltzman & Kelso, 1987). A very important point to be added here is that coordination in human perceptuo-motor skills deals with information, not only forces (Kelso, 1995; Warren, 2006). Firstly, perception and movement are tightly intermingled in behavioural coordination, and perception is all about information. Secondly the couplings between body parts, sensory systems, or neuronal populations, key to give rise and sustain coordination, are not only mechanical, but often only informational.

Discovering the laws underlying coordination in human behavior is made difficult because it embraces very many components, interactions often non linear between then, and an activity that evolves at different spatial and temporal scales (joints, muscles, neurons population, cells, ions channels, genes expression). Hence practically a large variety of measures, coordinates frames, can be used to probe a system's behaviour. Thankfully, patterned behavioral activity can be described by few collective variables that summarises the coordination. Note that similar types of reduction of dimensionality by considering macroscopic variables, patterns, is applied, for the purpose of theoretical modelling or constrained by the limitations of empirical measurement, at lower scales. In the former case one can think about modelling brain activity using neural mass or neural field approaches, and for the latter using surface electrode measurements to get a macroscopic measurement of muscular activation.

Such collective variables have been found generically close to a qualitative change (bifurcation) and modelled by use of the adiabatic elimination procedure, according to the slaving principle, a physical principle which contains a set of mathematical techniques, which corresponds to the physical intuition that "long-living variables enslave short-living variables" (Haken et al., 1985; Haken, 1988; Kelso, 1995). Basically when approaching a bifurcation the variables which will bifurcate relax more and more slowly, while the more stable variables relax fast. Consequently the slowly (relaxing) variables are ruling the game. Why? They contain most of the information about what the system is currently doing. At their time scale the stable variables are almost surely constant (hence the use of "adiabatic eliminination"), hence their evolution contains no information. These special variables express the low dimensional character of the overall organization. The dynamics of coordination patterns are described by dynamical systems, they are mapped onto attractors of collective variables in phase space, the stability of which is considered crucial, in that it pulls (attracts) the different elements (e.g., limbs, muscular synergies,…) of the system into the preferred overall mode of operation given task goal and other constraints. In such a complex system, composed of many elements, evolving at various spatial and temporal scales, behaving in a nonlinear way and with nonlinear interactions, it is not possible to deduce the patterns' behaviour from the behaviour of isolated components. The patterns dynamics, their stability and change, is not a trivial outcome of the behaviour of the components, the role played by the cooperation between the components is essential. A derivation of the system's behaviour from first principles or lower scales, be it mechanical laws, neural networks dynamics, neurophysiological processes, can be envisioned but remains a great scientific challenge. Phenomenological but physiologically plausible models can be reached, intermediate between microscopic processes and completely phenomenological models.

A level of abstraction

The principles of coordination dynamics, the basic laws of coordination, are defined with a level of abstractness. Agreeing with several attempts in the study of motor control, one here seeks lawful and invariant properties of behavior (e.g., motor equivalence, the Lambda model, the two third power law, Fitt's law, the minimum jerk model, the uncontrolled manifold theory). In this vein, basic and universal properties provided models of entrainment phenomena with non- linear systems of coupled oscillators found applications in several functional settings, recruiting very diverse components and couplings, ranging from mechanical to perceptual constraints (e.g., bimanual coordination, posture, locomotion, synchronization between individuals). The abstract focus is again emphasized when one aims at defining classes of qualitatively distinct behaviors (Huys et al., 2008; Jirsa and Kelso, 2005), considering the topological structure of dynamics of functionally relevant behavioral variables (fixed point attractors, limit cycles attractors, torus, chaotic attractors, excitable behavior, relaxation oscillations, separatrices between basin of attractions), and not the details of the underlying substrates. However this focus at an abstract level is dictated first by the behavior of the systems under consideration itself, which display a range of universal properties implemented in various structures and substrates, and by the difficulty to link different scales of description.

Noise induces variability

Coordination patterns have to resist unavoidable internal and external perturbations, hence must possess stability. Stability measures tell the therapist or the researcher what is /are the functionally relevant (stable) pattern(s). When the dynamic pattern is characterized by a fixed point, like in synchronization dynamics for instance, stability can be quantified in stationary behaviour by the variance of the fixed point, or by applying a small perturbation. For instance relative timing between rhythmically moving limbs proved a stable variable in that after a phasic perturbation, mechanical (Scholtz & Kelso, 1989; Scholtz Kelso & Schöner, 1987), or perceptual perturbation (Yamanishi, Kawato & Suzuki, 1979; Bardy et al., 2002), relative timing returns to its value before the perturbation was applied. In a pointing task, like the classical Fitts' task, stability confers essential functional properties: robustness against finite random perturbations inherent to the system, and thus the interplay between stability and random fluctuations may explain in principle the observed spatial variability in physical space. Note that from the early developments in the mathematical theory of dynamical systems the focus on stability was justified by the necessity to ensure persistence, robustness, and "observability" against fluctuations unavoidable in any real physical- natural system (Andronov et al., 1966, see page xviii of the Introduction chapter). Without random contributions no variability is allowed, the deterministic part of the models being perfectly regular. Human goal-directed behavior is characterized by multistability: different stable patterns are available for a fixed set of conditions and parameters. Two main types of variability are considered, one arising from the functional redundancy characterizing the movement, visible in the co-variation of the components of the effector system (e.g., joints), and one arising from the noise inherent to any biological system.

In the dynamical formulation of human coordination, the presence of variability has been modelled in a standard way by adding "dynamical" Gaussian delta correlated noise to the equations of motion (Schöner et al., 1986), under the hypothesis that underlying fluctuations evolve from more microscopic level. The assumptions underlying the use of random forces for modelling are an evolution at a faster time scale, with smaller amplitude than the actual functional behavior. In synergetics (Haken, 1988), the random forces are produced by smaller components. This is the assumptions used in applications to biology. The main idea, by taking this option, is to describe all the "structure", the correlations and order present in the behaviour, by the deterministic part of the model, hence the addition of random contributions that are in contrast uncorrelated. These assumptions are generally in agreement with the measurement of variability at much faster time scales than the adaptive movement, mainly found in various elements in the nervous system (Calvin and Stevens, 1967; Stein et al., 2005), ranging from ions channels opening at cortical synapses to ensemble neurons population. Recently various origins of noise acting on genes expression have been proposed, which may act in the nervous system, creating variability in the number of proteins produced from a specific gene (Isaacs et al., 2003; Raser et al., 2005): Intra cellular thermal motion of molecules, termed "thermal noise", noise in biochemical reactions, and chemical liaisons between molecules, sometimes termed "quantal noise". The microscopic size of the smallest of these biological so-called components make then sensitive to thermodynamic fluctuations, their enormous number may justify the use for approximation of Gaussian uncorrelated noise (the central limit theorem). Note that noise is also present at other levels involved in motor control and trajectory formation, for instance in muscles and in the environment. This modelling led to new quantitative predictions, deduced by these assumptions and based on Langevin and Fokker-Planck formulations[2] (Schöner et al., 1986), verified experimentally in phase transition phenomena (Bardy et al., 2002; Kelso et al., 1986; Schmidt et al., 1990).

A study of the interplay between determinist and random components requires gathering knowledge about the type of dynamics, of attractors' landscape, characterizing the system. Measures of stability which are defined rigorously for certain type of attractors cannot be generalized to others. For instance the variance of the relative phase in bimanual coordination (Schöner et al., 1986), or effector-environment events (Kelso et al., 1990) is an operational measure of stability of a fixed point attractor. Under specific assumptions this variable is explicitly related to relaxation time, and under the fluctuation-dissipation theorem (Gardiner, 1990) and the Wiener- Khinchine theorem[3] to the power spectrum line width and the autocorrelation function, which can also be estimated from empirical data. In the very illustrative case of bimanual coordination, and one of the most evolved stochastic modelling to date, two tasks are achieved (Schöner et al, 1986): 1) studying the local stationary solutions, and 2) studying the transients. In the first case the non linear equation describing the dynamics of the relative phase, a fixed point dynamics with two attractors, in phase and anti phase, at low rate of motion, is linearized around each fixed point. This local approximation enables to write down the linear Langevin equation and relate stability, amplitude of noise, mean, standard deviation, the power spectrum and autocorrelation function. The autocorrelation is interesting, as it tells you something about the dynamics, here the influence of the value of the variable at time (t) on the value at time (t + $\tau$). An expression was obtained for each fixed point, leading to the exact expression of the stationary distribution of the relative phase from the Fokker Planck equation, then of the mean and the variance. The expressions for the local stationary distributions (probability density distributions) can be related to an Ornstein- Uhlenbeck process, one of the most well known diffusion stochastic process, defined by a negative linear deterministic component (linear damping), corresponding to the so-called drift term, forced by additive uncorrelated noise, giving rise in the Fokker Planck equation to the so-called diffusion term. In this emblematic case, when stability decreases, meaning the absolute value of the so-called negative damping decreases, the variance increases, the line width of the power spectrum increase, and the correlations, given by the autocorrelation function increase. These systematic changes predicted by the analysis of the model are the product of the interplay between noise and the stability provided by the determinist component. This model led to new predictions, for instance when the antiphase pattern of coordination approaches a loss of stability critical fluctuations should arise, which were checked experimentally (Kelso et al., 1986). A complementary analysis was done to compute the distribution of the time of the transients between the two stable states (switching time), using the concept of mean first passage time, for which the agreement with empirical data found later was very good (Kelso et al., 1987). The rather drastic assumptions and the local linearization gave excellent predictions for stationary fluctuations when approaching switching behavior. Recall that it was assumed that the process under study is of the Markov type, that was, the probability to get the value of the relative phase at the current time is completely defined by the probability to get the relative phase in the previous time step. Stated differently, the probability density depends only on its present value and not on its history, meaning the coordination is considered of short range memory. The second assumption was that noise is additive and uncorrelated. These assumptions were conservative in some sense; after all, it consisted in approximating the stochastic dynamics by the simplest model, much in the classical idea of truncating at lower order terms an expansion of a more complex model, which is a typical first approach in physics which brought about many successes, like enabling Jean Perrin to obtain the Avogadro number from the diffusion coefficient formulated by Einstein in 1905 (Gardiner, 1990). However this approach may fail when approaching the bifurcation, as noted by Schöner et al. (1986). In this approximation the first two "statistical descriptor" of the probability distribution, the first moments, mean and variance, totally describe this distribution, for it is approximated by a Gaussian distribution (remember however the state is captured by an angle variable, the phase difference, which cannot be considered strictly as Gaussian); this is of course again guided by the tenet that higher moments capture more and more rare events, meaning only the tails of the distribution are then concerned. To get other features, for instance long range correlations, non linear drift and multiplicative noise may be necessary to consider (Franck et al., 2006). Bottom line the idea was to relate the stochastic process of interest to a known stochastic process, for each one knows that exact expressions for quantities that can be estimated from empirical data can be derived. In this case the stochastic process is Gaussian, and stationary solutions, inphase and antiphase, can be obtained from the model for a range of rate of movement.

Variance, and also coefficient of variation, have been considered in other perceptual-motor problems as indexes of stability without prior identification of the attractors, meaning of the states of the system studied. One example of such generalization is made in the study of postural processes by analysing the center of pressure trajectory. This is a misleading generalization from the insights gathered by the (stochastic) dynamical system approach outlined above, and from the much older assumption that the average trajectory reveals the underlying pattern, while the standard deviation between sets of individual trajectories and the average trajectory is considered to represent the degree to which the trajectories converge to the template, or the stability of the movement. In fact, with long enough data samples, the deterministic and the random components of the stochastic dynamics can be identified from data, under the Markov assumption, using the Kramer Moyals expansion, which requires estimating the conditional probability distribution which gives the probability to find the system in state x' at time t +τ, given a previous state x at time t (Friedrich and Peinke, 1997; see application to human movement in van Mourik et al., 2005).

To conclude this part, in the present framework, if no redundancy is present, and without random contributions: No variability is allowed except when chaos is present, the deterministic part of the systems giving solution being perfectly regular. Some difficulties can arise however because, in a fully deterministic system without any random contribution, quasi-periodic regime, deviation from purely harmonic oscillations, can give rise, to relatively irregular behaviour, hence possessing some type of *variability* (Bergé, Pommeau, & Vidal, 1984). Variability in movement was once attributed to quasi-periodic deterministic behaviours, when power spectra of time series of experimental data showed up with several unexpected harmonics (Schmidt & Turvey, 1993; but see Riley and Turvey, 2002 and Richardson et al., 2007). However it has later shown that the estimated variability was simply an artefact produced by the choice of coordinates used for the analysis (Fuchs and Kelso, 1994; Fuchs et al., 1996). It seems however that this "mysterious" variability, of another type than the one dealt with by classic Langevin and Fokker Planck frameworks, is still inspiring some. Very many confusions are still usual in the so called "motor control" literature about the status and origin of variability, and its relation to stability. The latter being sometimes ignored, or misunderstood. We saw that it doesn't make sense to discuss the relation between stability and variability without a clear definition of the problem being analyzed, in particular without identifying the regime in which the system is. As a further illustration, stability is sometimes associated with simple order. This may be intuitive, but it is wrong. Take the Kuramoto theory and model of collective synchronization. This model contains a stable incoherent regime , which proved difficult to study (See Strogatz 2000 for a presentation), and also partially synchronized coexistent states termed "chimeras". Therefore "stable" does not equate simple ordered organization. A more difficult problem is posed when variability is attributed to a chaotic behaviour. Here random fluctuations and deterministic mixing of trajectories are closer to each others in many respects. To the best of our knowledge, a clear cut indication of a deterministic chaotic property of behavior has not been demonstrated, and in this eventuality, its distinction with stochastic processes may be a (difficult) issue (See a response in favor to stochastic processes in Ramdani et al., 2011).

**Notes**

(1) Several approaches are advancing our understanding. Following the tenet that the CNS behaves as if minimizing uncertainty, and this comes in various developments (Friston, 2010; Harris and Wolpert, 1998; Hasson et al., 2016). Noteworthy are also the developments in control theory (Todorov and Jordan, 2002), and earlier use of the theory of communication (Schmidt et al., 1979).

(2) Fokker Planck equation and minimization of free energy have been formally related for the Brownian motion of a particle in a force a field, like the Ornstein- Uhlenbeck process (Jordan et al., 1998). The connections between entropy reduction, stochastic dynamics, and Shannon entropy have been studied (see with examples including biological coordination of movements, Haken, 1988).

(3) The Wiener Khinchine theorem shows that the power spectrum is the Fourier transform of the autocorrelation (see Bergé, Pommeau, & Vidal, 1984).

**Acknowledgements:** The author is indebted to Scott Kelso and Viktor Jirsa for their contributions to the interdisciplinary PhD training program proposed at the Center for complex systems and Brain Sciences, Florida Atlantic University.

References

Andronov, A. A., Vitt, A. A., & Khaikin, S. E. (1966). Theory of oscillators. Oxford, NY: Pergamon Press/Dover Publications.

Bergé, P., Pommeau, Y., Vidal, C. (1986). Order within Chaos. New York: Wiley-interscience.

Frank, T. D., Friedrich, R., & Beek, P. J. (2006). Stochastic order parameter equation of isometric force production revealed by drift-diffusion estimates. Physical Review E, 74, 051905.

Friedrich R, Peinke J, Renner C (1997) How to quantify deterministic and random influences on the statistics of the foreign exchange market. Phys Rev Lett 84:5224–5227.

Friston, K. (2010). The free-energy principle: a unified brain theory? Nature Reviews Neuroscience, 11(2), 127-138.

Fuchs, A., Jirsa, V.K., Haken, H., Kelso, J.A.S. (1996). Extending the HKB model of coordinated movement to oscillators with different eigenfrequencies. Human Movement Science, 74, 21-30.

Jirsa, V., Kelso, J.A.S. (2005) The excitator as a minimal model for the coordination dynamics of discrete and rhythmic movement generation. J Mot Behav 37(1):35–51.

Jordan, R., Kinderlehrer, D., & Otto, F. (1998). The variational formulation of the Fokker--Planck equation. SIAM journal on mathematical analysis, 29(1), 1-17.

Haken, H. (1988). Information and self-organization: A macroscopic approach to complex systems. Springer

Haken, H., Kelso, J.A.S., Bunz, H. (1985). A theoretical model of phase transitions in human bimanual coordination. Biological Cybernetics, 51, 347-56.

Harris, C. M., & Wolpert, D. M. (1998). Signal-dependent noise determines motor planning. Nature, 394(6695), 780-784.

Hasson, C. J., Zhang, Z., Abe, M. O., & Sternad, D. (2016). Neuromotor Noise Is Malleable by Amplifying Perceived Errors. PLoS Comput Biol, 12(8), e1005044.

Holmes, P. (2005). Ninety plus thirty years of non linear dynamics: Less is more and more is different. International Journal of Bifurcation and Chaos, 15, 2703-2716.

Huys, R., Studenka, B. E., Rheaume, N. L., Zelaznik, H. N. & Jirsa, V. K. (2008). Distinct timing mechanisms produce discrete and continuous movements. PLoS Comput Biol, 4, e1000061.

Isaacs, F.J. and Hasty, J. and Cantor, C.R. and Collins, JJ (2003). Prediction and measurement of an autoregulatory genetic module, Proceedings of the National Academy of Sciences, 100, 7714-7719.

Kelso, J.A.S. (1995). Dynamic patterns: the self-organization of brain and behavior. MIT Press, Cambridge.

Kelso, J.A.S., Scholz, J.P., Schöner, G. (1986). Non equilibrium phase transitions in coordinated biological motion: critical fluctuations. Physical Letters A, 118, 279-284.

Ramdani, S., Seigle, B., Varoqui, D., Bouchara, F., Blain, H., & Bernard, P. L. (2011). Characterizing the dynamics of postural sway in humans using smoothness and regularity measures. Annals of biomedical engineering, 39, 161-171.

Raser, JM, O'Shea, EK, 2005. Noise in genes expression: Origins, consequences, and control. Science, 309, 2010-2013.

Riley, M.A., Santana, M.V., Turvey, M.T. (2001) Deterministic variability and stability in detuned bimanual rhythmic coordination. Human Movement Science, 20, 343–369.

Riley, M.A., Turvey, M.T. (2002) Variability of determinism in motor behavior. Journal of Motor Behavior, 34, 99–125.

Saltzman, E., Kelso, J.A.S. (1987). Skilled actions: a task dynamic approach. Psychological Review, 94, 84–106.

Schmidt, R.C., Shaw, B.K., Turvey, M.T. (1993). Coupling dynamics in interlimb coordination. Journal of Experimental Psychology: Human Perception and Performance, 19, 397-415.

Schmidt, R. A., Zelaznik, H., Hawkins, B., Frank, J. S., & Quinn Jr, J. T. (1979). Motor-output variability: a theory for the accuracy of rapid motor acts. Psychological review, 86, 415.

Schöner, G., Haken, H., Kelso, J.A.S. (1986). A stochastic model of phase transitions in human hand movement. Biological Cybernetics, 53, 247-257.

Strogatz, S. H. (2000). From Kuramoto to Crawford: exploring the onset of synchronization in populations of coupled oscillators. Physica D: Nonlinear Phenomena, 143, 1-20.

Strogatz, S.H. (2004). Nonlinear Dynamics and Chaos. With Applications to Physics, Biology, Chemistry, and Engineering. Cambridge, MA.

Todorov, E., & Jordan, M. I. (2002). A minimal intervention principle for coordinated movement. In Advances in neural information processing systems (pp. 27-34).

Warren, W.H. (2006). The dynamics of perception and action. Psychological Review, 113, 358-389.

van Kampen, N.G. (1997) Stochastic Processes in Physics and Chemistry. Elsevier, Amsterdam.

van Mourik, A.M. ,Daffertshofer, A., Beek, P.J. (2006). Estimating Kramers–Moyal coefficients in short and non-stationary data sets. Physical Letter A, 351, 13–17.